# Metal-Organic Framework-Derived Sugarcoated Haws-like AgNWs/ZIF-8/Pd for Plasmon-Promoted Photocatalytic Hydrogenation

*Shuoren Li[†], Rui Wu[†], Xingxing Meng[†], Leilei Diao[†], Jing Wang[†], Chuanping Li\*[†,‡]*

[†]Anhui Province Key Laboratory of Functional Coordinated Complexes for Materials Chemistry and Application, School of Chemical and Environmental Engineering, Anhui Polytechnic University, Wuhu 241000, P.R. China.

[‡]State Key Laboratory of Electroanalytical Chemistry, Changchun Institute of Applied Chemistry, Chinese Academy of Sciences, 5625 Renmin Street, Changchun 130022, P. R. China.

ABSTRACT

The conversion of biomass-derived materials into value-added products via photocatalysis holds significant promise in driving the development of renewable resources. However, since catalytic processes often require high temperatures and pressures, most catalysts are quite difficult to meet the requirements of high selectivity and high activity simultaneously. Herein, a plasmon-promoted photocatalyst, integrating Ag nanowires with metal-organic frameworks (MOFs)-confined Pd nanoparticles, is rationally designed to afford AgNWs/ZIF-8/Pd which achieve highly selective and efficient catalytic hydrogenation toward 2(5H)-Furanone under mild conditions. The plasmonic AgNWs/ZIF-8/Pd exhibited much stronger photocatalytic activity and higher selectivity compared with AgNWS/ZIF-8 and ZIF-8/Pd. The enhanced activity can be attributed to the synergistic coupling between Pd nanoparticles and AgNWs, and the possible reaction mechanism is proposed. This work provides a new approach to constructing efficient photocatalysts and offers new insights into the understanding of the influence of the plasmonic effect on photocatalytic hydrogenation reactions.

**INTRODUCTION**

Depleting fossil resources and escalating global energy challenges underscore the critical need to develop sustainable resources.[1-4] Biomass, as a vast and renewable energy source, holds great potential for addressing these concerns. It can be efficiently converted into fuels and value-added chemicals through chemical or biological processes.[5-7] In the last decades, many approaches have been investigated to achieve the high-efficient conversion of biomass materials. Among many different approaches, photocatalysis is considered to be very promising due to its mild reaction conditions and environmentally friendly reaction process.[8-12] I In this context, the development of efficient photocatalysts is of great importance. 2(5H)-Furanone, derived from renewable biomass sources, has gained considerable attention as a platform chemical with versatile reactivity and potential applications in pharmaceuticals, agrochemicals, and polymer synthesis.[13, 14] Considering its chemical structure characteristics, 2(5H)-furanone can be used to efficiently synthesize γ-butyrolactone (GBL) and 1,4-butanediol (BDO). Currently, GBL is primarily produced from fossil feedstock by dehydrocyclization of BDO and hydrogenation of maleic anhydride. Due to the shortage of fossil resources, it has become more attractive to produce GBL using renewable materials. However, the hydrogenation of 2(5H)-Furanone usually requires high temperature and $H_2$ pressure,[15, 16] and most catalysts fail to meet this dual requirement of high chemoselectivity and activity simultaneously due to difficulties in cleavage of stable C=O bonds. Obviously, the design of a suitable catalyst for the hydrogenation of 2(5H)-Furanone toward γ-butyrrolactone with high chemoselectivity and efficiency under mild conditions is highly desirable.

Metal-organic frameworks (MOFs) have garnered significant attention in recent years due to their remarkable electronic, optical, and catalytic properties. The unique characteristics of MOFs, including their chemical stability, high specific surface area, and tunable chemical properties, make them particularly attractive for catalytic applications.[17-20] Among the various MOFs, ZIF-8 stands out as a prominent example due to its exceptional potential for photocatalytic applications.[21, 22] This is primarily attributed to the presence of organic functional groups within its structure, which contribute to its inherent photocatalytic activity. However, due to some inherent properties of ZIF-8, such as low charge transfer efficiency and wide band gap, there is still a gap in performance with conventional photocatalytic materials.[22-24] To overcome the hindrance, researchers have taken various approaches. For example, the construction of ZIF-8 and $TiO_2$ composites reduced the band gap to 2.98 eV compared to pure ZIF-8.[25] Carbonization of MOFs to porous carbon in an inert or reducing atmosphere[26, 27] and combination with other nanomaterials to improve photoconversion. Recent research shows that the localized surface plasmon resonance (LSPR) effect derived from the noble metal could not only expand the visible light absorption of ZIF-8 but also further reduce the band gap.[28] Therefore, an effective way to enhance the photocatalytic ability of ZIF-8 is to combine light-trapping noble metal nanomaterials with ZIF-8. Silver nanowires (AgNWs), as a typical noble metal nanowire, are perfect candidates to improve the disadvantages of ZIF-8 compared to oxidation-prone copper nanowires and expensive gold nanowires.[29] In addition, the AgNWs exhibit strong and broad LSPR absorption characteristics in the wavelength range of 330-900 nm compared with silver nanoparticles.[30]

Herein, a novel plasmon-enhanced MOF-based catalyst AgNWs/ZIF-8/Pd is prepared via in-situ growing ZIF-8 on AgNWs to form sugarcoated haws-like nanostructures and subsequently anchoring small-size Pd nanoparticles due to the nanoconfinement of ZIF-8. AgNWs as a noble metal based plasmonic catalyst, displays strong LSPR in the visible light region. ZIF-8 was an effective candidate in enriching $H_2$ in a nanoscale microenvironment. On the other hand, the plasmonic AgNWs could inject plasmon-induced electrons to the Pd active site and finally result in the high-efficient catalytic activity of the AgNWs/ZIF-8/Pd. As expected, the AgNWs/ZIF-8/Pd presents exceptionally higher catalytic activities for the hydrogenation under visible-light irradiation and are far superior to those of AgNWs/ZIF-8, AgNWs/Pd and ZIF-8/Pd. The photocurrent measurement further confirmed the fast separation and transfer of plasmon-induced charge carriers within AgNWs/ZIF-8/Pd.

EXPERIMENTAL SECTION

**Chemicals**

$AgNO_3$, ethanol, methol, ethylene glycol, $Zn(NO_3)_2·6H_2O$, NaCl were purchased from Sinopharm Reagent Co. (Beijing, China). Palladium(II) acetate, 2(5H)-Furanone was purchased from Adamas-bata (Shanghai, China). $HAuCl_4$, 2-methylimidazole, Polyvinyl Pyrrolidone (PVP, Mw~58000) from Aladdin Reagent Company (Shanghai, China). All the chemicals were used without further purification.

**Synthesis of AgCl**

This synthesis is performed in the dark due to the photosensitivity of AgCl. $AgNO_3$ (3mL, 0.5M) is mixed with NaCl (3mL, 1M) under stirring for 1 min. The precipitate is separated from the supernatant, washed once with DI water and dried under vacuum. It is noteworthy that the use of freshly prepared AgCl is essential for the successful preparation of AgNWs.[31]

**Synthesis of AgNWs**

In a typical synthesis, 0.68g of PVP (Mw~58000) is dissolved into 40mL of ethylene glycol in a 100mL flask and brought to 160°C. Once the solution has reached a stable temperature, 50mg of freshly prepared AgCl is added all at once. After 3 min, 0.22g of $AgNO_3$ are added all at once. The reaction mixture is left under stirring at 160°C for 24 min. After cooling, wash once with deionized water and disperse into methanol solution and reserve.

**Synthesis of AgNWs/ZIF-8**

0.40833g of 2-methylimidazole was dissolved in 10mL of AgNWs methanol solution. Under magnetic stirring, 10mL of methanol solution containing 0.35g $Zn(NO_3)_2·6H_2O$ was added all at once and reacted for 30 min. The products were collected by centrifugation and dried under vacuum for use. The resulting products were named AgNWs/ZIF-8.

**Synthesis of AgNWs/ZIF-8/Pd**

5.6mg palladium acetate was dissolved in 150mL methanol at 30°C water bath under magnetic stirring. After 15 min stirring, 0.15g AgNWs/ZIF-8 was dispersed into the above solution, and continuously stirred for 30 min, then the product was separated by centrifugation and washed with ethanol three times noted as AgNWs/ZIF-8/Pd.

**Electrochemical experiments**

The PEC experiments were conducted with CHI 660E in an electrolyte containing 0.5M $NaSO_4$ (pH=5.6) by using a 300 W xenon lamp as the illumination light. The working, reference and counter electrodes were AgNWs/ZIF-8/Pd, saturated calomel electrode (SCE), and platinum wire, respectively.

**Structure and Property Characterization**

The morphologies were recorded by a scanning electron microscope (SEM, Hitachi S4800) and transmission electron microscopy (TEM, FEI TalosF200x). X-Ray Diffraction (XRD) was obtained by using a D8 ADVANCE instrument (Bruker, Germany). The X-ray photoelectron spectroscopy (XPS) was characterized by using an XPS analysis (Thermo ESCALAB 250). UV–vis spectra were performed on a UV–Vis spectrophotometer (Shimadzu UV-3600). The products were analyzed by using Aglint GC-8890 chromatograph.

**Catalytic test**

The catalytic activity of AgNWs/ZIF-8/Pd was examined by the hydrogenation of 2(5H)-Furanone to γ-butyrrolactone. The catalytic reaction conditions: 5mg catalyst, 0.0594mmol substrate, 0.03mmol NH3BH3, 10mL ethanol reaction time 1h, reaction temperature 25℃, under visible light irradiation.

## RESULTS AND DISCUSSION

**Synthesis and Characterization of AgNWs/ZIF-8/Pd.**

The typical synthesis process of AgNWs/ZIF-8/Pd is depicted in Figure. 1a. First, the AgNWs were prepared in this study according to the previously reported method.[31] As illustrated in Figure. S1a, Ag exhibited a wire-like morphology with a length of ~ 50 μm. The prepared AgNWs were dispersed in methanol solution, to which 2-methylimidazole was added and mixed, and then AgNWs/ZIF-8 was prepared by adding an equal volume of methanol solution of zinc nitrate and stirring at room temperature (Figure. S1b). Pd nanoparticles was anchored on AgNWs/ZIF-8 by using a mild reduction method in the presence of methanol as reducing agent. After deposition of Pd nanoparticles, uniformly distributed nanoparticles were observed between AgNWs and ZIF-8 as shown in Figure S2, and their average size was about 6.01 nm (Figure S3). Benefitting from the nanoconfined effect of the pore/channel structures in ZIF-8, the reaction precursors could be retained inside the well-confined pores and precipitated via a "ship-in-a-bottle" route.[32] Figure. 1b showed the SEM images of the AgNWs/ZIF-8/Pd, The HRTEM images (Figure 1d) presented the lattice spaces of 0.235nm, 0.281nm and 0.241nm, matched well with the (1 1 1), (2 4 4) and (0 1 0) planes of Ag, ZIF-8 and Pd, respectively. This result was consistent with the elemental mapping images shown in Figure. 1f-i, and the sugarcoated haws-like nanostructures is expected to promote the enhancement of the catalytic activity of hot electron injection during the catalytic hydrogenation reaction, as well as the capture of reactants and $H_2$.

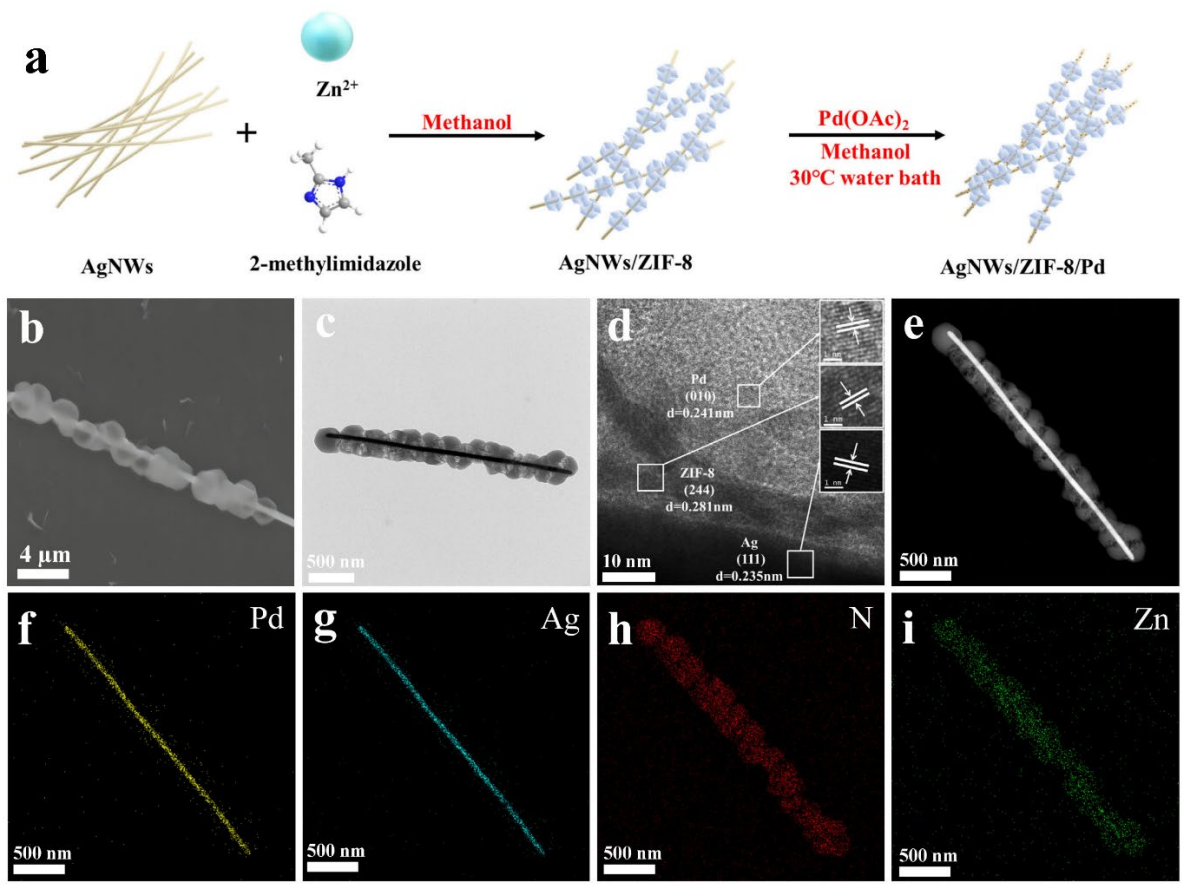

**Figure. 1** (a) The synthesis procedure of AgNWs/ZIF-8/Pd, (b) SEM image, (c) TEM, (d) HRTEM, (e) HAADF, (f-i) elemental mapping of AgNWs/ZIF-8/Pd.

Figure. S4 showed the X-ray diffraction (XRD) patterns of ZIF-8, AgNWs/ZIF-8 and AgNWs/ZIF-8/Pd. The obvious crystal peaks at 2θ=7.43° and 12.49° were well-matched with the standard card of ZIF-8 (JCPDS No. 00-062-1030).[33] while the diffraction peaks of AgNWs were identical to that of the AgNWs (JCPDS No. 96-150-9147).[31] Owing to the particularly low amount of Pd nanoparticles, no typical peaks of Pd appeared in the XRD patterns. However, from the XPS spectra (Figure. 2c and S5), the elements of Pd were confirmed, which strongly identified the successful anchoring of Pd nanoparticles. From Figure. 2a, b, S6 a, b AgNWs/ZIF-8/Pd exhibited strong signals of elemental Zn, Ag, O and N. Compared to the standard AgNWs sample, the Ag 3d spectrum has a slight blue shift in the two peaks at 367.7eV (Ag 3d$_{5/2}$) and 373.7eV (Ag 3d$_{3/2}$).[34] indicating that the creation of obvious space charge region at the interface of AgNWs and ZIF-8, which lead to electron transportation. To further confirm this electron transportation, the Zn 2p peaks high-resolution scan spectra were shown in Figure. 2a. The 2p spectrum of Zn consists of two peaks at 1021.8eV and 1044.8eV, corresponding to Zn 2p$_{3/2}$ and Zn 2p$_{1/2}$, respectively, compatible with $Zn^{2+}$.[35] Compared to pure ZIF-8, the composite showed no significant blue shift or red shift. A significant red shift was observed in the high-resolution scan spectra of Pd 3d.

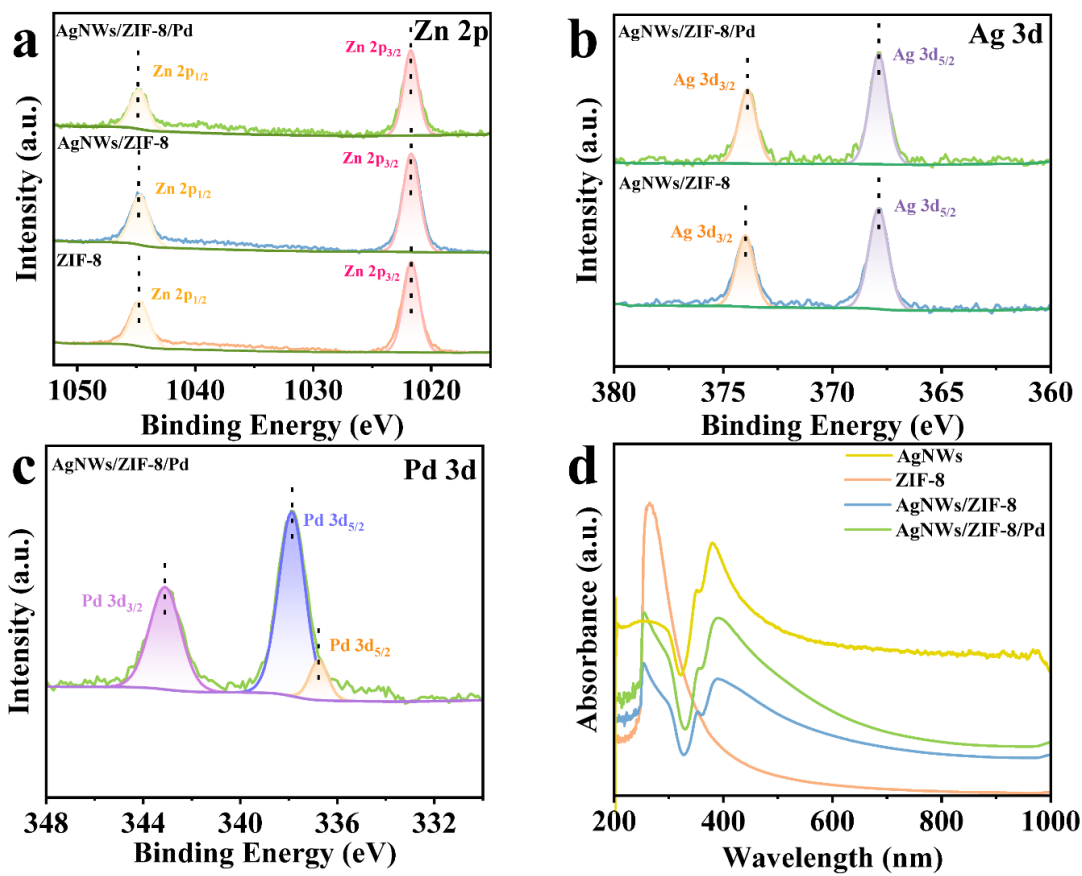

**Figure. 2** High-resolution XPS spectra of (a)Zn 2p, (b) Ag 3d and (c) Pd 3d, (d) UV-Vis spectra of the sample.

The tight binding of AgNWs to ZIF-8 is achieved by in situ growth, and Pd exists between AgNWs and ZIF-8 via mild reduction, and this structure can provide a possible charge transport pathway for site-based photocatalysis, thus improving the efficiency of photocatalysis. The successful preparation of ZIF-8 in AgNWs/ZIF-8 and AgNWs/ZIF-8/Pd can be well illustrated by the N1s and O1s spectra shown in Figure S6a,b. Figure 2d shows the UV-Vis spectra of different samples, AgNWs/ZIF-8/Pd exhibit strong and broad LSPR absorption characteristics in the wavelength range of 330-900 nm. All the results strongly evidenced the successful preparation of AgNWs/ZIF-8/Pd.

**Catalytic hydrogenation of 2(5H)-Furanone over AgNWs/ZIF-8/Pd**

The catalytic activity of AgNWs/ZIF-8/Pd was examined by the catalytic reaction of 2(5H)-Furanone. The catalytic reaction was conducted in the sealed reactor containing ammonia borane ($NH_3BH_3$) as $H_2$ source both under visible light irradiation and in the dark. In the absence of catalyst addition, the γ-butyrrolactone were formed under both light and dark conditions, indicating that the decomposition of ammonia borane promotes the reaction. As shown in Figure 3, AgNWs/Pd play a dominant role in the catalytic reaction, while the yields of the other samples are below 20% under visible light irradiation and dark conditions. According to previous work, it is known that Pd, Pt and Ru nanoparticle-loaded catalysts are widely used in catalytic hydrogenation reactions due to their high affinity for $H_2$,[16, 32] and Pd nanoparticles play an important role in this work. This is consistent with the results shown in Figure 3. The reaction yield catalyzed by AgNWs/Pd under visible light irradiation was 57.3% while that catalyzed by AgNWs was only 8.3%. Similarly, the yields catalyzed with ZIF-8/Pd and ZIF-8 under visible light irradiation were 15.7% and 8.3%, respectively. The large difference between AgNWs/Pd and ZIF-8/Pd may be attributed to the localized surface plasmon resonance of AgNWs with strong plasmonic coupling and hot electron injection from AgNWs to Pd nanoparticles. Furthermore, the yields of AgNWs/ZIF-8/Pd under visible light irradiation was 87.5%, which is the highest among the photocatalysts. This enhancement was attributed to the plasmonic effect of AgNWs in the generation of hot electrons and the formation of Pd active sites upon absorption of visible light. Furthermore, the excellent $H_2$ enrichment ability of ZIF-8, derived from the nanoconfined effect of the pore/channel structures, played a great role in increasing the reaction activity while improving the reaction selectivity.

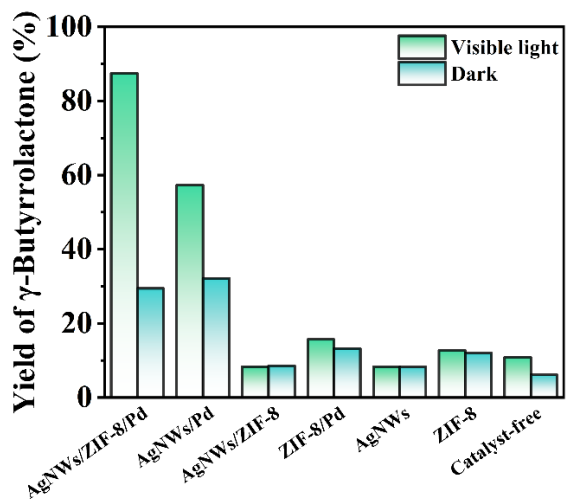

**Figure. 3** The comparison of catalytic hydrogenation reaction activity in 60 min with NH3BH3 as a reducing agent over different samples at 25 °C, both under visible light irradiation and in the dark.

As shown in Figure S7, the conversion reached 32.4% under the condition of visible light irradiation without catalyst addition and reached less than 50% conversion for all samples except AgNWs/ZIF-8/Pd and AgNWs/Pd, which indicated the large amount of side reactions present in this reaction. According to previous studies, 2(5H)-Furanone undergoes polymerization reactions under this reaction condition.[36] Moreover, the 91.6% conversion and 87.5% yield obtained for AgNWs/ZIF-8/Pd under the conditions of visible light irradiation can effectively illustrate the high efficiency and selectivity of this structure. To further investigate the kinetic factors of the reaction, we explored the effect of reaction time on this catalytic reaction, and the yields and conversions showed an increasing trend with increasing reaction time under the same reaction conditions (Figure. S8a, b). In addition, the activation energy of the catalytic reaction was calculated by Arrhenius plots, the activation energies (Ea) as 14.86 KJ/mol (Figure. S9).

**Photoelectrochemical test**

The photoelectrochemical tests was studied to offer important hints for plasmon-induced hot electron transfer and recombination. As shown in Figure 4a, AgNWs exhibited a negligible photocurrent response under intermittent illumination, and this result was in remarkable agreement with its catalytic performance. In contrast, ZIF-8, ZIF-8/Pd, AgNWs/ZIF-8 and AgNWs/ZIF-8/Pd exhibit a fast photocurrent response under intermittent light irradiation at the glassy carbon electrode, and this response is stable and can be well reproduced over several on-off cycles. It can be also observed that AgNWs/ZIF-8/Pd shows a considerably high photocurrent intensity due to the transfer of hot electrons generated from visible light irradiation on AgNWs to Pd through the Schottky junction. The enhancement of photocurrent intensity from AgNWs/ZIF-8 to AgNWS/ZIF-8/Pd can well illustrate the above point and this result is in good agreement with its catalytic performance. From the finite-difference time-domain (FDTD) simulation result shown in Figure 4b, the electromagnetic (EM) field intensity at the interface of AgNWs and Pd or ZIF-8 was significantly enhanced and larger hot-spot areas were generated compared to the AgNWs/ZIF-8. Such an enhanced EM field has been proven to improve the electron-hole separation.

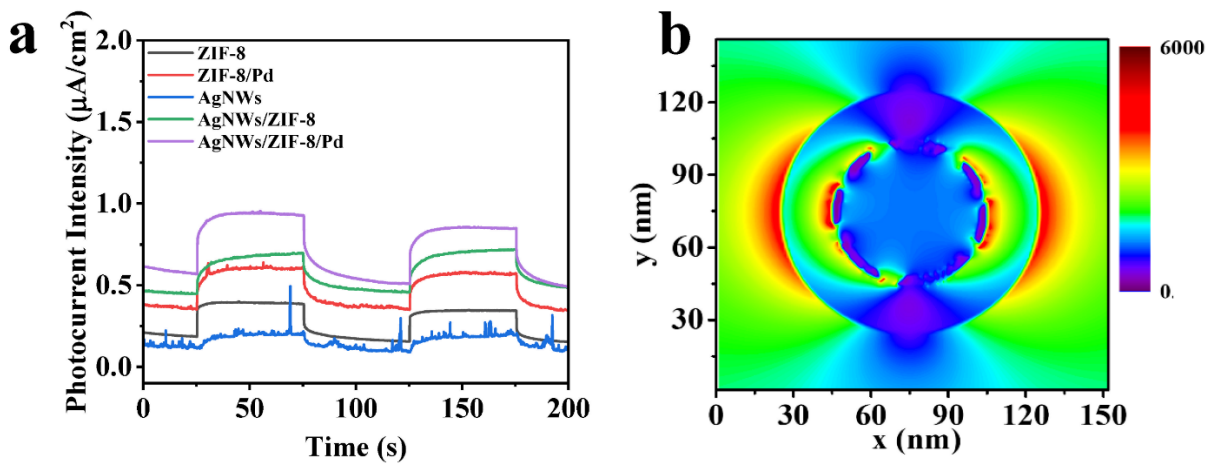

**Figure. 4** (a) Photocurrent responses of different samples with or without visible light irradiation, (b) FDTD simulation

**Possible reaction mechanism**

Based on the above discussion, the possible mechanism of catalytic hydrogenation by plasmonic AgNWS/ZIF-8/Pd under visible light irradiation was investigated. As shown in Figure. 5, hot electrons with high energy are excited through plasmonic effect under the irradiation of visible light. Then plasmon-induced high-energy hot electrons can be injected directly into the Pd nanoparticles, resulted in the formation of the electron-rich Pd nanoparticles.[32] The catalytic hydrogenation reaction with the participation of Pd nanoparticles involves the adsorption of reactants onto Pd nanoparticles to form activated complexes, followed by the dissociation of stable C=O bonds and the formation of C-H bonds in the presence of $H_2$ source. However, the catalytic activity of the Pd-based catalysts is hindered by the weak binding of the reactants to the noble metal active sites and the difficulty in breaking the stable C=O bonds. In this study, the intrinsic catalytic activity of Pd-catalyzed 2(5H)-Furanone hydrogenation is promoted due to the high density of electrons in Pd nanoparticles, while the existence of ZIF-8 can effectively trap the reactants and $H_2$, and the process can be well facilitated by the $H_2$ generated from $NH_3BH_3$ dehydrogenation.[37, 38] Meanwhile, the plasmon-induced positive charge may react with $H_2O$ to form hydroxyl radicals, which then react with $NH_3BH_3$ to produce $H_2$.[39] Finally, the electrons consumed during the reaction are continuously replenished by dehydrogenation of $NH_3BH_3$. In summary, the reasonable design of AgNWs/ZIF-8/Pd composite structure enables the hot electrons to be injected into the Pd nanoparticles, and the reactants and $H_2$ are well enriched around the Pd active site, which effectively improves the catalytic hydrogenation reaction activity and selectivity.

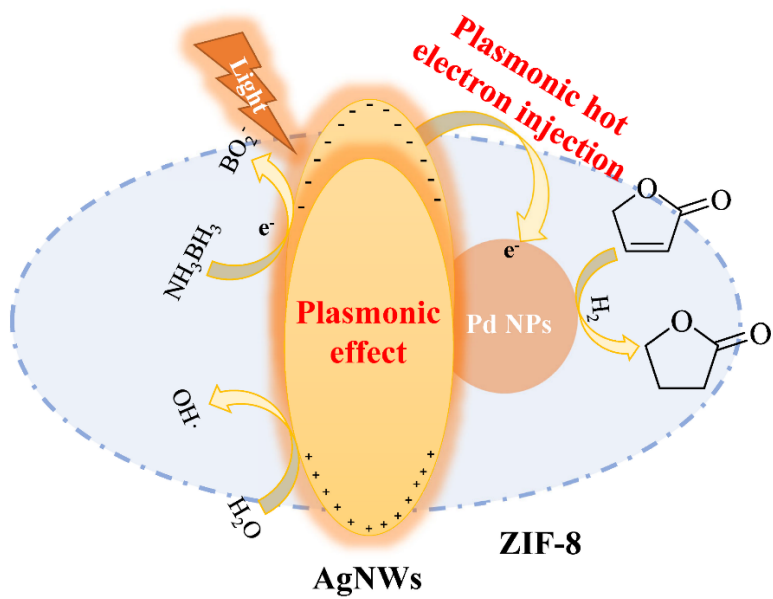

**Figure. 5** Schematic diagram of the possible reaction mechanism for the hydrogenation of 2(5H)-Furanone over plasmonic AgNWs/ZIF-8/Pd

**SUMMARY**

In summary, we developed a plasmon-enhanced sugarcoated haws-like photocatalysts (AgNWs/ZIF-8/Pd) to achieve a high-efficient catalytic activity toward hydrogenation of 2(5H)-Furanone. The electron-rich Pd nanoparticles, derived from the plasmonic coupling and hot electron injection effect of AgNWs, and the excellent $H_2$ enrichment ability of ZIF-8 played great role in boosting the catalytic activity. This work provides new insights into the understanding of LSPR on photocatalytic hydrogenation reactions and provides an effective approach for the designing of high-efficient photocatalysts.

## ASSOCIATED CONTENT

The Supporting Information is available free of charge.

## AUTHOR INFORMATION

**Corresponding Author**

**\*Chuanping Li**- *School of Chemical and Environmental Engineering, Anhui Polytechnic University, Wuhu 241000, Anhui, China;* Email: licp@ahpu.edu.cn

**Notes**

The authors declare no competing financial interest.

## ACKNOWLEDGMENT

This work was supported by the China National Natural Science Fund (No. 22004002), the Natural Science Foundation of Anhui Province (No. 2008085QB80), the Open Funds of the State Key Laboratory of Electroanalytical Chemistry (SKLEAC202103), the Open Fund of Anhui Laboratory of Functional Coordinated Complexes for Materials Chemistry and Application (LFCCMCA-10) and Research Start-up Fund of Anhui Polytechnic University (No. 2019YQQ016).